%% file: sample-sigconf.tex
\documentclass[sigconf]{acmart}
\usepackage{hyperref}
\usepackage{subfigure}
\usepackage{enumerate}
\usepackage{graphicx}
\usepackage{url}
\usepackage{amsmath}
\usepackage{booktabs}
\usepackage{enumerate}
\usepackage{ulem}
\usepackage{multirow}
\usepackage{makecell}
\usepackage{titlesec}
\usepackage{balance}
\usepackage[ruled,vlined]{algorithm2e}
\newcommand{\m}{DualIPW}

\AtBeginDocument{%
  }

\copyrightyear{2025}
\acmYear{2025}
\setcopyright{cc}
\setcctype{by}
\acmConference[WWW Companion '25]{Companion Proceedings of the ACM Web Conference 2025}{April 28-May 2, 2025}{Sydney, NSW, Australia} \acmBooktitle{Companion Proceedings of the ACM Web Conference 2025 (WWW Companion '25), April 28-May 2, 2025, Sydney, NSW, Australia} \acmDOI{10.1145/3701716.3715458} \acmISBN{979-8-4007-1331-6/2025/04}

\begin{document}


\settopmatter{printacmref=true}
\linepenalty=1000

\title[Unbiased Learning to Rank with Query-Level Click Propensity Estimation]{Unbiased Learning to Rank with Query-Level Click Propensity Estimation: Beyond Pointwise Observation and Relevance}
\author{Lulu Yu}
\affiliation{%
\institution{CAS Key Lab of Network Data Science and Technology, ICT, CAS}
  \institution{University of Chinese Academy of Sciences}
  \city{Beijing}
  \country{China}}
\email{yululu23s@ict.ac.cn}

\author{Keping Bi}
\authornote{Corresponding author.}
\affiliation{%
\institution{CAS Key Lab of Network Data Science and Technology, ICT, CAS}
  \institution{University of Chinese Academy of Sciences}
  \city{Beijing}
  \country{China}}
\email{bikeping@ict.ac.cn}

\author{Jiafeng Guo}
\affiliation{%
\institution{CAS Key Lab of Network Data Science and Technology, ICT, CAS}
  \institution{University of Chinese Academy of Sciences}
  \city{Beijing}
  \country{China}}
\email{guojiafeng@ict.ac.cn}

\author{Shihao Liu}
\author{Dawei Yin}
\affiliation{
\institution{Baidu Inc.}
\city{Beijing}
\country{China}
}
\email{liushihao@baidu.com}
\email{yindawei@acm.org}

\author{Xueqi Cheng}
\affiliation{%
\institution{CAS Key Lab of Network Data Science and Technology, ICT, CAS}
  \institution{University of Chinese Academy of Sciences}
  \city{Beijing}
  \country{China}}
\email{cxq@ict.ac.cn}

\renewcommand{\shortauthors}{Lulu Yu et al.}




\input{sections/abs}

\maketitle

\input{sections/intro}
\input{sections/preliminary}
\input{sections/method}

\input{sections/exp}

\input{sections/conclusion}
\balance
\begin{acks}
This work was funded by the National Natural Science Foundation of China (NSFC) under Grants No. 62302486, the Innovation Project of ICT CAS under Grants No. E361140, the CAS Special Research Assistant Funding Project, the project under Grants No. JCKY2022130C039, the Strategic Priority Research Program of the CAS under Grants No. XDB0680102, and the NSFC Grant No. 62441229.
\end{acks}
\linepenalty=1000
\normalem
\bibliographystyle{ACM-Reference-Format}
\bibliography{ref}
\end{document}

%% file: sections/abs.tex
\begin{abstract}
Most existing unbiased learning-to-rank (ULTR) approaches are based on the user examination hypothesis, which assumes that users will click a result only if it is both relevant and observed (typically modeled by position). However, in real-world scenarios, users often click only one or two results after examining multiple relevant options, due to limited patience or because their information needs have already been satisfied. Motivated by this, we propose a query-level click propensity model to capture the probability that users will click on different result lists, allowing for non-zero probabilities that users may not click on an observed relevant result. We hypothesize that this propensity increases when more potentially relevant results are present, and refer to this user behavior as relevance saturation bias. Our method introduces a Dual Inverse Propensity Weighting (\m) mechanism---combining query-level and position-level IPW---to address both relevance saturation and position bias. Through theoretical derivation, we prove that \m~can learn an unbiased ranking model. Experiments on the real-world Baidu-ULTR dataset demonstrate that our approach significantly outperforms state-of-the-art ULTR baselines. The code and dataset information can be found at \url{https://github.com/Trustworthy-Information-Access/DualIPW}.

\end{abstract}
\keywords{Unbiased Learning to Rank, Position Bias, Relevance Saturation Bias}

%% file: sections/intro.tex
\section{Introduction}
Most existing unbiased learning-to-rank (ULTR) studies focus on mitigating position bias \cite{joachims2017unbiased,ai2018unbiased,luo2024unbiased,zhang2023towards} and are based on position-based click modeling \cite{richardson2007predicting}. Position-based modeling (PBM) assumes that users click a result when they examine the result and it is relevant. The observation probability (i.e., click propensity) depends on the position of the result on the search result page (SERP). Grounded on PBM, a major thread of ULTR research is to estimate the propensity of each position on the SERP and conduct inverse propensity weighting (IPW) on the click \cite{joachims2017unbiased,ai2018unbiased,luo2024unbiased}. Among them, a popular approach is the dual learning algorithm (DLA) \cite{ai2018unbiased} that adaptively learns relevance and click propensity by IPW and inverse relevance weighting simultaneously.


Despite its efficacy on synthetic click data, DLA does not perform as well on the Baidu-ULTR dataset \cite{zou2022large}, real-world large-scale click data \cite{hager2024unbiased}. There are several potential reasons for this. First, users may have more complex behaviors PBM cannot capture \cite{zou2022large}. Second, search engines have strong logging policies that rank high-relevance results to higher positions, leading to inaccurate propensity estimation from confounded relevance and observation \cite{hager2024unbiased}. Third, severe false negative issues exist in the result list. Most results are relevant but users only click one or two of them \cite{yu2023cir}.

\enlargethispage{\baselineskip}
Existing work has made efforts to address these challenges. To capture complex user behaviors, \citet{chen2021adapting} propose the Interactional Observation-Based Model (IOBM), which estimates a position's click propensity by incorporating clicks from other positions in the result list. To mitigate issues arising from strong logging policies, \citet{zhang2023towards} introduce gradient reversal (GradRev) and observation dropout (Drop) to disentangle relevance and observation. \citet{luo2024unbiased} propose an unconfounded propensity estimation (UPE) approach with backdoor adjustment. To address the false negative problem, \citet{wang2021non} argue that non-clicks with higher observation probability are more likely to be true negatives, and therefore assign higher weights to pairs formed by a click and such non-clicks. However, this approach implies that higher-position results are more likely to be true negatives, which contradicts the fact that strong logging policies tend to rank more relevant results higher. Since most of these methods are evaluated on synthetic data, their performance on real-world click data remains unclear.

Targeting effective ULTR with real-world click data, we challenge the examination hypothesis in PBM. In our analysis of the Baidu-ULTR dataset, we find that among query sessions with clicks, only about 1.2\% involve more than two clicks, approximately two-thirds have a single click, and one-third have exactly two clicks. Given users' limited patience and the fact that a single relevant result typically satisfies their information needs, we hypothesize that users may have a non-zero probability of not clicking a relevant result, even after examining it. As the number of potentially relevant results in a list increases, users are less likely to leave the session without clicking. This behavior reflects a query-level click propensity influenced by the quality of the ranking list, which we refer to as relevance saturation bias.

Accordingly, we propose a Dual Inverse Propensity Weighting mechanism (\m) that incorporates both query-level and position-level click propensity. When the query-level propensity is larger, the ranking list of the query has a smaller inverse weight, which reduces the importance of sessions with potentially more false negatives. Hence, inaccurate contrastive learning between clicks and non-clicks will be mitigated. We theoretically prove that \m~can learn an unbiased ranking model. We compare \m~with state-of-the-art baselines, including GradRev, Drop, UPE, and IOBM, on the Baidu-ULTR dataset and demonstrate significant improvements over them. 

%% file: sections/preliminary.tex
\section{Preliminary}
\subsection{Dual Learning Algorithm (DLA)}
DLA \cite{ai2018unbiased} is an effective method for mitigating position bias based on the user examination hypothesis \cite{richardson2007predicting}: 
\begin{equation}
p(c_d=1)=p(r_d=1)\cdot p(o_d=1)\text{,}
\end{equation}
where $c_d$, $r_d$, and $o_d$ represent whether a document d is clicked, relevant, and observed.
DLA treats the estimation of relevance and propensity as a dual problem. Given a query $q$ and its document list $\pi_q$, DLA alternates the learning of an unbiased ranking model $f$ and an unbiased propensity model $g$ by inverse propensity weighting (IPW) and inverse relevance weighting (IRW):
\begin{small}
\begin{equation}
l_{IPW}(f,q)=\!\!\!\!\!\!\!\!\sum_{d\in \pi_q, c_d=1}\!\!\!\!\frac{l(f(x_d),c_d)}{p(o_d=1)}\text{, }l_{IRW}(g,q)=\!\!\!\!\!\!\!\!\sum_{d\in \pi_q, c_d=1}\!\!\!\!\frac{l(g(k_d),c_d)}{p(r_d=1)}\text{,}
\end{equation}\end{small}
where $k_d$, $x_d$, and $l$ denote the document at position $k$, features of $d$ regarding $q$, and the loss function, respectively.
\subsection{Pilot Study}
\label{data_ref}
\subsubsection{The Relationship Between Click Positions and Relevance Saturation (RS) Bias:} A more severe relevance saturation (RS) bias leads to a higher number of false negatives (relevant but non-clicked results), which in turn can degrade model performance due to inaccurate labels. To explore how RS bias varies across query sessions, we divide the Baidu-ULTR query sessions into subsets and evaluate models trained on each subset. Since we have limited query-level information, we use click positions within a query as the partition criterion. For simplicity, we focus on single-click sessions, which account for two-thirds of the sessions with clicks. 

Fig. \ref{single} shows the average nDCG@10 for models trained on single-click sessions at different positions, using results from 5 random seeds. We focus primarily on the first four positions, as they constitute about 92\% of all the single-click sessions. The results show that sessions with single clicks at positions 1 and 4 perform significantly better than those at positions 2 and 3, suggesting less severe RS bias and fewer false negatives. Notably, the model trained on single-click sessions at position 4, despite having much less training data, still performs relatively well. This could be attributed to the way search results are displayed on mobile devices, where the first three results typically fit on the screen. Users may scroll down and click the fourth result when the first three are irrelevant, while clicking the second or third result does not imply that the results before them are irrelevant.

\subsubsection{Features for Estimating Relevance Saturation Bias:} We then explore the features that can distinguish RS bias for queries with different click sequences. We attempt to estimate result quality, correlated to RS bias, with a surrogate ranking model. We train the model with DLA using sessions with clicks and use it to predict relevance scores. We tried several unsupervised query performance prediction (QPP) methods to estimate result quality but they are not distinctive regarding the single-click sessions at different positions. We then explore and identify another discriminative feature. For a click sequence, $\mathbf{cs}$, with a single click, e.g., ($1,0,\dots,0$), we calculate the proportions of the maximal score of the list occurring at each position and obtain a distribution of the maximal-score positions, denoted as $D_{mp}^{\mathbf{cs}}$ according to: 
\begin{small}
\begin{equation}
\label{dms}
    D_{mp}^{\mathbf{cs}}=(\frac{\sum_{q\in Q,\mathbf{cs}^q=\mathbf{cs}}\mathbb{I}(i ==\arg\max_{j \in \{1,\dots, 10\}} s_j)}{\sum_{q\in Q}\mathbb{I}(\mathbf{cs}^q==\mathbf{cs})})_{i=1,\dots,10}\text{,}
\end{equation}\end{small}where $\mathbf{cs}^q$ indicates the click sequence of query $q$, $s_j$ represents the relevance score of the document at position $j$ predicted by the surrogate model, and $\mathbb{I}$ is an indicator function. Fig. \ref{whole} (left) also illustrates how $D_{mp}^{\mathbf{cs}}$ is calculated. 

\begin{figure}[!t]
\centering
\hspace{-1mm}
\subfigure{
\includegraphics[width=0.23\textwidth]{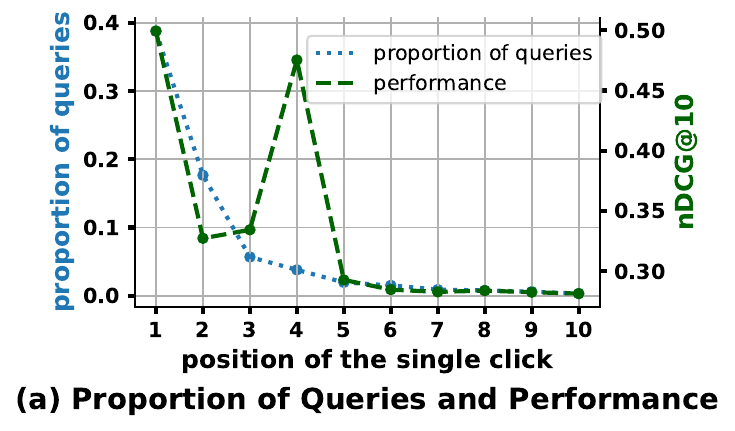}
\label{single}}\subfigure{
\includegraphics[width=0.24\textwidth]{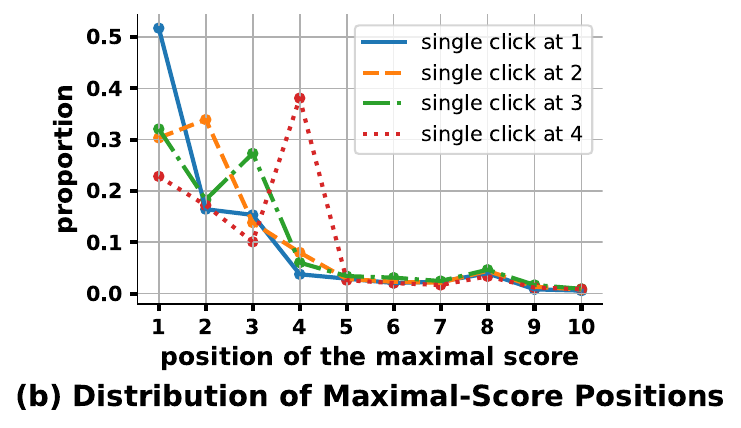}
\label{single_feature}}
\caption{(a) The distribution of single-click sessions at each position and performance of models trained by 10 single-click groups. (b) The distribution of maximal-score positions of single-click sessions at positions 1, 2, 3, and 4.}
\end{figure}

We extracted \( D_{mp}^{\mathbf{cs}} \) for single-click sessions at the first four positions and present the four distributions in Fig. \ref{single_feature}. By comparing Fig. \ref{single} and Fig. \ref{single_feature}, we observe that the more consistent \( D_{mp}^{\mathbf{cs}} \) is with \( \mathbf{cs} \), the better the model's performance—indicating less severe relevance saturation bias. This suggests that we can use the divergence of \( D_{mp}^{\mathbf{cs}} \) from its original \( \mathbf{cs} \) to estimate relevance saturation bias across different click sequences.

%% file: sections/method.tex
\section{Methodology}
\subsection{Dual Click Propensity Hypothesis}
\label{theory}
We assume that queries with more relevant documents are more likely to receive clicks, although only one or two. More relevant results lead to more severe relevance saturation bias. We refer to the probability of a query $q$ receiving clicks as the query-level click propensity (denoted as $cp_q$). Based on this assumption, we propose a new click hypothesis, where $c_q$ represents whether a query $q$ receives clicks:\looseness=-1
\begin{equation}
\label{hypothesis}
p(c_d=1)=p(o_d=1)\cdot p(r_d=1)\cdot p(c_q=1)\text{.}
\end{equation}
To mitigate relevance saturation bias and position bias, we propose Dual Inverse Propensity Weighting (\m) with inverse query-level and position-level propensity to compute loss:
\begin{equation}
\mathcal{L}_{\m}(f)=\int_{q\in Q} \sum_{d\in \pi_q, c_d=1}\frac{ l(f(x_d),c_d)\text{d}p(q)}{p(o_d=1)\cdot p(c_q=1)}\text{.}
\end{equation}
where $q$ is drawn from $Q$ according to $q\sim p(q)$.
Theoretically, $\mathcal{L}_{\m}(f)$ is an unbiased estimation of the global loss $\mathcal{L}(f)$, where $\mathbf{o}^q=(o_d)_{d\in \pi_q}$, and $\mathbf{c}=(c_q)_{q\in Q}$,
\begin{footnotesize}
\label{proof}
\begin{align}
\mathbb{E}_{\mathbf{o}^q,\mathbf{c}}[\mathcal{L}_{\m}(f)]
&=\mathbb{E}_{\mathbf{c}}[\int_{q\in Q}\mathbb{E}_{\mathbf{o}^q}[\sum_{\text{d}\in \pi_q, c_d=1}\frac{ l(f(x_d),c_d)\text{d}p(q)}{p(o_d=1)\cdot p(c_q=1)}]]\nonumber\\
&=\mathbb{E}_{\mathbf{c}}[\int_{q\in Q}\mathbb{E}_{\mathbf{o}^q}[\sum_{d\in \pi_q, r_d=1}\frac{o_d\cdot c_q\cdot l(f(x_d),r_d)\text{d}p(q)}{p(o_d=1)\cdot p(c_q=1)}]]\nonumber\\
&=\mathbb{E}_{\mathbf{c}}[\int_{q\in Q}\sum_{d\in \pi_q, r_d=1}\frac{c_q\cdot l(f(x_d),r_d)\text{d}p(q)}{p(c_q=1)}]\\
&=\int_{q\in Q} \sum_{d\in \pi_q,r_d=1} l(f(x_d),r_d)\text{d}p(q)=\mathcal{L}(f)\nonumber\text{.}
\end{align}\end{footnotesize}The proof consists of two levels: 1) Inner Level: mitigating position bias at the position level within each query with inverse position-level propensity weighting (\textbf{PL-IPW}), and 2) Outer Level: mitigating relevance saturation bias at the query level across different queries with inverse query-level click propensity weighting (\textbf{QL-IPW}). Since they are isolated, we can similarly prove that the \m~can learn an unbiased propensity model as shown in \cite{ai2018unbiased}.

\subsection{Query-Level Click Propensity Estimation}
\begin{figure}[t]
    \centering
    \includegraphics[width=1\linewidth]{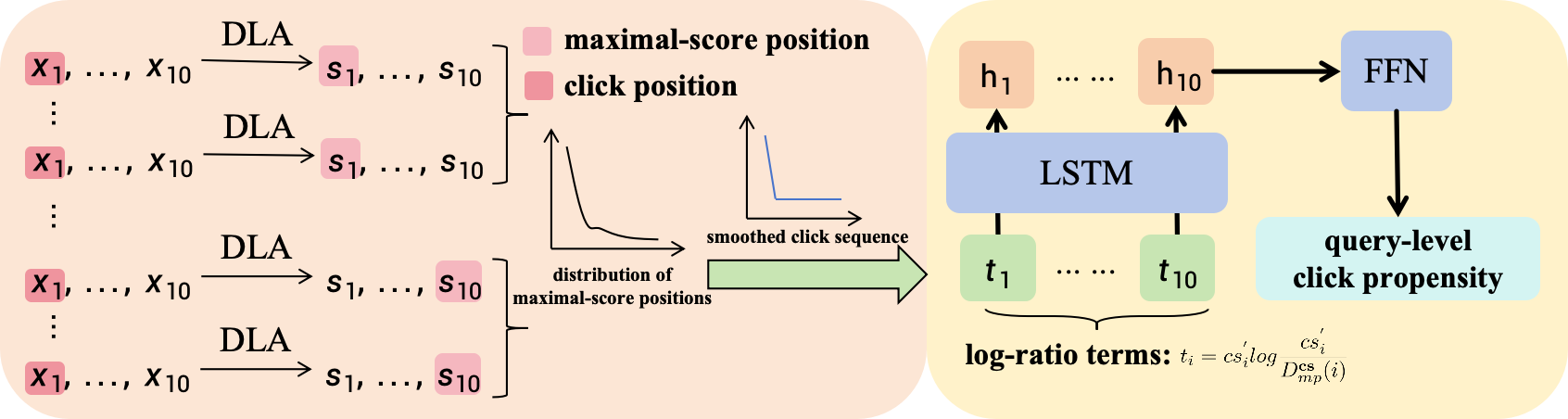}
    \caption{Query-level click propensity estimation.}
    \label{whole}
\end{figure}

Section \ref{data_ref} shows that the divergence of $D_{mp}^{\mathbf{cs}}$ from its original click sequence $\mathbf{cs}$ can indicate relevance saturation bias, so we use the divergence between them to estimate the query-level click propensity. Since the click sequence of $n$ positions has $2^n$ possible combinations, we need to calculate $2^n$ distributions for all possible sequences, which is cumbersome. For simplicity, we treat multi-click sequences as combinations of multiple single-click sequences. Thus, given a click sequence $\mathbf{cs}$, its query-level click propensity can be calculated as the average propensity of single-click sequences, where the subscript $i$ represents the single-click sequence with a click at position i,
\begin{equation}
cp_{\mathbf{cs}}=\frac{\sum_{i=1,cs_i=1}^{10}cp_{\mathbf{cs}^i}}{\sum_{i=1}^{10}\mathbb{I}(cs_i==1)}\text{.}
\end{equation}
Then we only need to estimate the click propensities of single-click sequences. 
Considering that the divergence at each position contributes differently to the final query-level click propensity, we use the log-ratio terms, ${(cs}_ilog\frac{{cs}_i}{D_{mp}^{\mathbf{cs}}(i)})_{i=1,\dots,10}$, as the listwise input. 
Then, as shown in Fig. \ref{whole} (right), we employ an LSTM model to encode the hidden representations of these log-ratio terms and use a feed-forward network (FFN) on the last hidden state of the sequence to predict the query-level click propensity. We use the last state since lower positions could be more important in reflecting relevance saturation bias. We smooth the original click sequence with softmax: $\mathbf{cs}'=\textrm{softmax}(\mathbf{cs}/\tau)$. Formally, the output of the query-level click propensity model for a specific click sequence $\mathbf{cs}$ is as follows:
\begin{equation}
\label{qw}
h(\mathbf{cs})\!=\!\textrm{FFN}(\textrm{LSTM}(t_1,\dots,t_{10})[-1])\text{, }t_i\!=\!{cs}_i'log\frac{{cs}'_i}{D_{mp}^{\mathbf{cs}}(i)}\text{,}
\end{equation}
where $h$ denotes the query-level click propensity model. The final predicted query-level click propensity of each single-click sequence is calibrated by the $\textrm{softmax}$ function as,
\begin{equation}
({cp}_{\mathbf{cs}^1},\dots,{cp}_{\mathbf{cs}^{10}})=\textrm{softmax}(h(\mathbf{cs}^1),\dots,h(\mathbf{cs}^{10}))\text{.}
\end{equation}
\subsection{Learning Algorithm}
In contrast to DLA which learns a position-level propensity model and a relevance model, \m~ has an additional query-level propensity model $h(\mathbf{cs})$ in Eq. \ref{qw} to learn. Since the query-level propensities use listwise information of a click sequence independent of LTR features, we update $h(\mathbf{cs})$ and relevance model $f$ simultaneously according to Eq. \ref{ipw}. To separate the effect of query-level and position-level propensity, we freeze $h(\mathbf{cs})$ during the position-level propensity model $g$ learning, in Eq. \ref{irw}:
\begin{small}
\begin{align}
    l(f,h,q)&=-\frac{cp_{\mathbf{cs}^1}}{cp_{\mathbf{cs}^q}}\cdot\sum_{d\in \pi_q,c_d=1}\frac{g(k_1)}{g(k_d)}\cdot \textrm{log}\frac{e^{f(x_d)}}{\sum_{d'\in \pi_q} e^{f(x_{d'})}}\text{,}\label{ipw} \\
    l(g,q)&=-\frac{cp_{\mathbf{cs}^1}}{cp_{\mathbf{cs}^q}}\cdot\sum_{d\in \pi_q,c_d=1}\frac{f(x_1)}{f(x_d)}\cdot \textrm{log}\frac{e^{g(k_d)}}{\sum_{d'\in \pi_q} e^{g(k_{d'})}}\text{.} \label{irw}
\end{align}
\end{small}

%% file: sections/exp.tex
\section{Experimental Setup}
\subsection{Dataset}
Baidu-ULTR dataset\footnote{\url{https://github.com/ChuXiaokai/baidu_ultr_dataset/}} provides search sessions with clicks collected from a Chinese search engine, Baidu, and a test set with human annotations. We train all the models with a subset of the click data used in the NTCIR-17 ULTRE-2 task \cite{niu2023overview}. Each query-document pair is represented by 14 scalar features, including term-based features (e.g., BM25) and the relevance score generated by the pre-trained BERT-based ranking model\footnote{\url{https://github.com/lixsh6/Tencent_wsdm_cup2023}}. After excluding sessions with fewer than 10 results or without clicks at the top 10 positions, 485,342 query sessions are kept for training. We split the test set and used 1,446 for validation and 5,201 for testing.
\subsection{Baselines}
To demonstrate the effectiveness and necessity of our proposed method, we compare the following methods. \textbf{BM25:} BM25 \cite{robertson1994some} ranks documents based on term matching. \textbf{Naive:} It directly uses raw click data to train a ranking model without debiasing. \textbf{IPW:} Inverse Propensity Weighting (IPW) \cite{joachims2017unbiased} uses the estimated propensity based on positions to re-weight clicks. \textbf{DLA:} Dual Learning Algorithm (DLA) \cite{ai2018unbiased} simultaneously learns an unbiased ranking model and an unbiased propensity model. \textbf{PRS:} Propensity Ratio Scoring \cite{wang2021non} assigns higher weights to pairs formed with a click and non-clicks with higher observation probability. \textbf{GradRev} and \textbf{Drop:} These two methods \cite{zhang2023towards} mitigate the negative confounding effect between relevance and position by unlearning the relevance in the propensity model. \textbf{UPE:} Unconfounded Propensity Estimation (UPE) \cite{luo2024unbiased} leverages backdoor adjustment to mitigate the over-estimation of propensities. \textbf{IOBM:} Interactional Obsevation-Based Model (IOBM) \cite{chen2021adapting} leverages click information besides positions to estimate propensities.

\subsection{Evaluation Metrics}
We use normalized Discounted Cumulative Gain (nDCG) and Expected Reciprocal Rank (ERR) at 1, 3, 5, and 10. Experimental results are averaged over 5 runs with different random seeds.

\subsection{Implementation Details} We established our baselines using ULTRA \cite{tran2021ultra}. The input features were initially projected to 64 dimensions, and the architecture of the ranking model is a Deep Neural Network (DNN) with three hidden layers of sizes [64, 32, 16]. For the LSTM block in the query-level click propensity model, we tuned the hidden size to $\{4,8,16\}$, and the number of hidden layers to $\{1,2\}$. We used the optimizer AdamW with a learning rate ranging from 2e-6 to 6e-6. We set the batch size as 30 (the number of queries) and trained 2 epochs. We fixed the size of the ranking list to 10 in the training stage.
\section{Experimental Results}
\subsection{Overall Performance}
From Tab. \ref{real_perform}, we observe: \textbf{1)} For real-world click data with more complex biases, estimating propensities related only to positions (IPW, DLA) does not outperform the Naive method. 
\textbf{2)} Methods that model complex user behaviors (i.e., IOBM) and disentangle observation and relevance for strong logging policies (i.e., UPE, GradRev, and Drop) are effective on real-world click data.  
\textbf{3)} PRS performs the worst, indicating that its assumption---non-clicked documents with higher observation probability are more likely to be true negatives---does not hold in real-world click data. 
\textbf{4)} Our \m~achieves significant improvements over the above best-performing method IOBM, validating the dual click propensity hypothesis in mitigating complex biases in real-world click data. 
\begin{table}[t]
\setlength{\tabcolsep}{0.2pt}
\footnotesize
 \centering
  \caption{Average performance from runs with 5 random seeds. \m, its ablations, and baselines are compared. The best overall and baseline performance is marked in \textbf{bold} and \uline{underlined}. Significant improvements over Naive, the best baseline, and significant degradations from \m~are marked with ``$*$'', ``$\dag$'', and ``$-$'' respectively ($p\leq0.05$).}
 \begin{tabular}{cllllllll}\toprule
   \multirow{2}{*}{Methods} & \multicolumn{4}{c}{nDCG@K} & \multicolumn{4}{c}{ERR@K}
    \\\cmidrule(r){2-5}\cmidrule{6-9}
             & K=1 & K=3 & K=5 & K=10 & K=1 & K=3 & K=5 & K=10\\\midrule
BM25&0.4139&0.4290&0.4445&0.4769&0.1395&0.2203&0.2480&0.2696\\\hline
Naive&0.4365&0.4518&0.4654&0.4948&0.1522&0.2359&0.2631&0.2838\\\hline
PRS&0.3382&0.3554&0.3713&0.4085&0.1215&0.1929&0.2185&0.2408\\\hline
IPW&0.4349&0.4492&0.4627&0.4925&0.1532&0.2365&0.2633&0.2839\\
DLA&${0.4392}^*$&0.4525&0.4660&0.4952&0.1527&0.2361&0.2633&0.2840\\\hline
GradRev&${0.4377}$&${0.4546}^*$&${0.4679}^*$&${0.4974}^*$&${0.1525}$&${0.2372}^*$&${0.2643}^*$&${0.2849}^*$\\
Drop&${0.4391}^*$&${0.4535}^*$&${0.4667}^*$&${0.4961}^*$&${0.1528}$&${0.2367}^*$&${0.2639}^*$&${0.2845}^*$\\
UPE&${0.4407}^*$&${0.4557}^*$&${0.4684}^*$&${0.4983}^*$&$\uline{0.1540}^*$&$\uline{0.2382}^*$&${0.2652}^*$&${0.2859}^*$\\\hline
IOBM&$\uline{0.4413}^*$&$\uline{0.4566}^*$&$\uline{0.4701}^*$&$\uline{0.5002}^*$&${0.1537}^*$&${0.2381}^*$&$\uline{0.2654}^*$&$\uline{0.2861}^*$\\\hline
\m&$\textbf{0.4429}^{*}$&$\textbf{0.4582}^{*\dag}$&$\textbf{0.4719}^{*\dag}$&$\textbf{0.5020}^{*\dag}$&$\textbf{0.1545}^{*}$&$\textbf{0.2391}^{*\dag}$&$\textbf{0.2664}^{*\dag}$&$\textbf{0.2871}^{*\dag}$\\
${-\textrm{PL-IPW}}$&${0.4391}^-$&${0.4556}^-$&${0.4691}^-$&${0.4990}^-$&${0.1540}$&${0.2383}$&${0.2654}^-$&${0.2861}^-$\\
${-\textrm{QL-IPW}}$&${0.4392}^-$&${0.4525}^-$&${0.4660}^-$&${0.4952}^-$&${0.1527}^-$&${0.2361}^-$&${0.2633}^-$&${0.2840}^-$
    \\\bottomrule
 \end{tabular}
 \label{real_perform}
\end{table}

\subsection{Ablation Study}
\m~comprises QL-IPW and PL-IPW. To validate the necessity of introducing both components, we conduct an ablation study, as shown in Tab. \ref{real_perform}. The PL-IPW here is essentially DLA. We can see that \m~significantly outperforms either component alone, confirming the importance of their combination.

\subsection{Fine-grained Comparison}
We further evaluate the performance of different ULTR methods on three levels of search frequencies. Since almost all metrics show similar trends, we take nDCG@10 as an example as shown in Fig. \ref{freq}. \m~demonstrates significant performance improvement on low-frequency queries, while the improvement on mid-frequency queries is marginal. On high-frequency queries, \m's performance is on par with methods other than IPW.

\subsection{Click Weight Analysis}
\begin{figure}[!t]
\centering
\subfigure{
\includegraphics[width=0.23\textwidth]{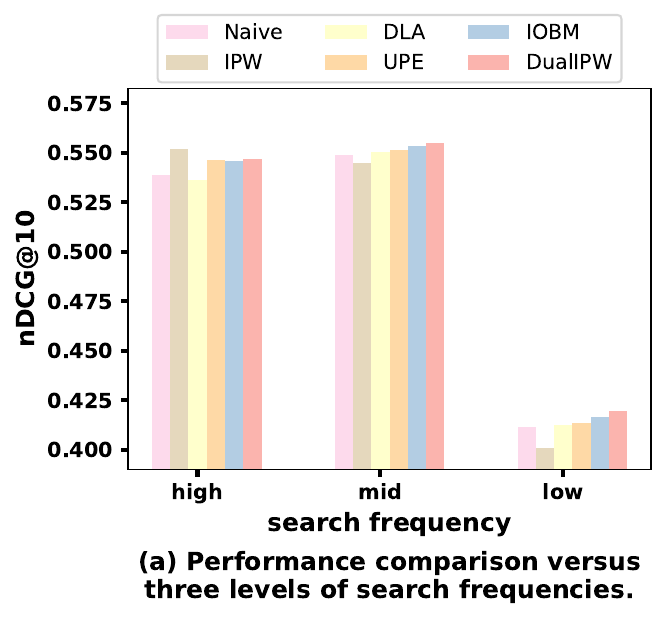}
\label{freq}}\subfigure{
\includegraphics[width=0.25\textwidth]{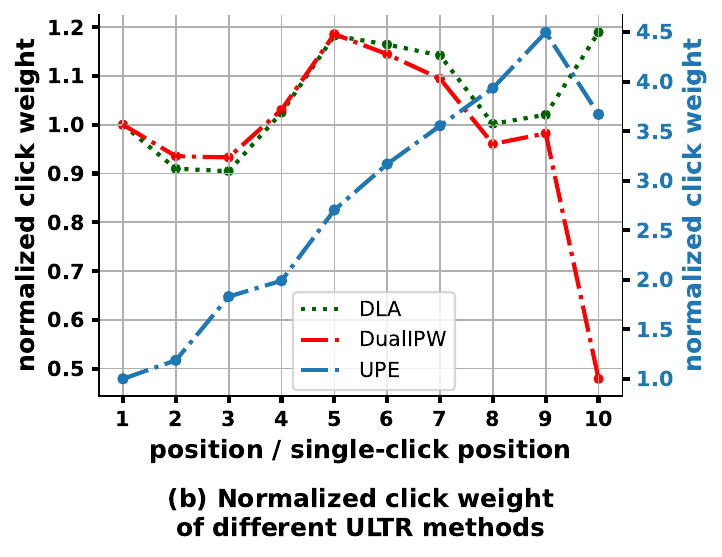}
\label{weight1}}
\caption{Fine-grained analysis and click weights.}
\end{figure}

We compare the click weight learned by different ULTR methods: IPW learned by DLA and UPE, and combined IPW learned by \m, as shown in Fig. \ref{weight1}. Since the PL-IPW learned by \m~and the IPW learned by DLA exhibit minimal differences, we omit the presentation of the former. Compared to the IPW learned on synthetic click data, which increases with position, the IPW learned by DLA on the real-world click data differs significantly, whereas UPE shows minimal difference. Compared to DLA, \m~adjusts the click weight by increasing it for clicks at the top ranks and decreasing it for lower ranks. Since queries with more relevant documents have higher click propensities, the click weight adjustment indicates that queries with a single click at lower ranks tend to have more relevant documents.

%% file: sections/conclusion.tex
\section{Conclusion}
In this work, we introduce relevance saturation bias, where queries with numerous relevant documents are more likely to receive clicks. We refer to a query's probability of receiving clicks as the query-level click propensity. To characterize relevance saturation bias, we propose a new click hypothesis: the click probability of a document is determined by its relevance, observation probability, and query-level click propensity. Based on this hypothesis, we propose the Dual Inverse Propensity Weighting (\m) method, which contains inverse query-level click propensity weighting and inverse position-level propensity weighting to alleviate relevance saturation bias and position bias, respectively. We demonstrate that leveraging \m~can learn an unbiased ranking model and achieve superior performance compared to existing ULTR methods on the Baidu-ULTR dataset. 

%% file: sample-sigconf.bbl

\begin{thebibliography}{13}


\ifx \showCODEN    \undefined \def \showCODEN     #1{\unskip}     \fi
\ifx \showISBNx    \undefined \def \showISBNx     #1{\unskip}     \fi
\ifx \showISBNxiii \undefined \def \showISBNxiii  #1{\unskip}     \fi
\ifx \showISSN     \undefined \def \showISSN      #1{\unskip}     \fi
\ifx \showLCCN     \undefined \def \showLCCN      #1{\unskip}     \fi
\ifx \shownote     \undefined \def \shownote      #1{#1}          \fi
\ifx \showarticletitle \undefined \def \showarticletitle #1{#1}   \fi
\ifx \showURL      \undefined \def \showURL       {\relax}        \fi
\providecommand\bibfield[2]{#2}
\providecommand\bibinfo[2]{#2}
\providecommand\natexlab[1]{#1}
\providecommand\showeprint[2][]{arXiv:#2}

\bibitem[Ai et~al\mbox{.}(2018)]%
        {ai2018unbiased}
\bibfield{author}{\bibinfo{person}{Qingyao Ai}, \bibinfo{person}{Keping Bi}, \bibinfo{person}{Cheng Luo}, \bibinfo{person}{Jiafeng Guo}, {and} \bibinfo{person}{W~Bruce Croft}.} \bibinfo{year}{2018}\natexlab{}.
\newblock \showarticletitle{Unbiased learning to rank with unbiased propensity estimation}. In \bibinfo{booktitle}{\emph{SIGIR'18}}. \bibinfo{pages}{385--394}.
\newblock


\bibitem[Chen et~al\mbox{.}(2021)]%
        {chen2021adapting}
\bibfield{author}{\bibinfo{person}{Mouxiang Chen}, \bibinfo{person}{Chenghao Liu}, \bibinfo{person}{Jianling Sun}, {and} \bibinfo{person}{Steven~CH Hoi}.} \bibinfo{year}{2021}\natexlab{}.
\newblock \showarticletitle{Adapting interactional observation embedding for counterfactual learning to rank}. In \bibinfo{booktitle}{\emph{SIGIR'21}}. \bibinfo{pages}{285--294}.
\newblock


\bibitem[Hager et~al\mbox{.}(2024)]%
        {hager2024unbiased}
\bibfield{author}{\bibinfo{person}{Philipp Hager}, \bibinfo{person}{Romain Deffayet}, \bibinfo{person}{Jean-Michel Renders}, \bibinfo{person}{Onno Zoeter}, {and} \bibinfo{person}{Maarten de Rijke}.} \bibinfo{year}{2024}\natexlab{}.
\newblock \showarticletitle{Unbiased Learning to Rank Meets Reality: Lessons from Baidu's Large-Scale Search Dataset}. In \bibinfo{booktitle}{\emph{SIGIR'24}}. \bibinfo{pages}{1546--1556}.
\newblock


\bibitem[Joachims et~al\mbox{.}(2017)]%
        {joachims2017unbiased}
\bibfield{author}{\bibinfo{person}{Thorsten Joachims}, \bibinfo{person}{Adith Swaminathan}, {and} \bibinfo{person}{Tobias Schnabel}.} \bibinfo{year}{2017}\natexlab{}.
\newblock \showarticletitle{Unbiased learning-to-rank with biased feedback}. In \bibinfo{booktitle}{\emph{WSDM'17}}. \bibinfo{pages}{781--789}.
\newblock


\bibitem[Luo et~al\mbox{.}(2024)]%
        {luo2024unbiased}
\bibfield{author}{\bibinfo{person}{Dan Luo}, \bibinfo{person}{Lixin Zou}, \bibinfo{person}{Qingyao Ai}, \bibinfo{person}{Zhiyu Chen}, \bibinfo{person}{Chenliang Li}, \bibinfo{person}{Dawei Yin}, {and} \bibinfo{person}{Brian~D Davison}.} \bibinfo{year}{2024}\natexlab{}.
\newblock \showarticletitle{Unbiased Learning-to-Rank Needs Unconfounded Propensity Estimation}. In \bibinfo{booktitle}{\emph{SIGIR'24}}. \bibinfo{pages}{1535--1545}.
\newblock


\bibitem[Niu et~al\mbox{.}(2023)]%
        {niu2023overview}
\bibfield{author}{\bibinfo{person}{Zechun Niu}, \bibinfo{person}{Jiaxin Mao}, \bibinfo{person}{Qingyao Ai}, \bibinfo{person}{Lixin Zou}, \bibinfo{person}{Shuaiqiang Wang}, {and} \bibinfo{person}{Dawei Yin}.} \bibinfo{year}{2023}\natexlab{}.
\newblock \showarticletitle{Overview of the NTCIR-17 Unbiased Learning to Rank Evaluation 2 (ULTRE-2) Task}.
\newblock \bibinfo{journal}{\emph{NTCIR'17}} (\bibinfo{year}{2023}).
\newblock


\bibitem[Richardson et~al\mbox{.}(2007)]%
        {richardson2007predicting}
\bibfield{author}{\bibinfo{person}{Matthew Richardson}, \bibinfo{person}{Ewa Dominowska}, {and} \bibinfo{person}{Robert Ragno}.} \bibinfo{year}{2007}\natexlab{}.
\newblock \showarticletitle{Predicting clicks: estimating the click-through rate for new ads}. In \bibinfo{booktitle}{\emph{WWW'07}}. \bibinfo{pages}{521--530}.
\newblock


\bibitem[Robertson and Walker(1994)]%
        {robertson1994some}
\bibfield{author}{\bibinfo{person}{Stephen~E Robertson} {and} \bibinfo{person}{Steve Walker}.} \bibinfo{year}{1994}\natexlab{}.
\newblock \showarticletitle{Some simple effective approximations to the 2-poisson model for probabilistic weighted retrieval}. In \bibinfo{booktitle}{\emph{SIGIR’94}}. Springer, \bibinfo{pages}{232--241}.
\newblock


\bibitem[Tran et~al\mbox{.}(2021)]%
        {tran2021ultra}
\bibfield{author}{\bibinfo{person}{Anh Tran}, \bibinfo{person}{Tao Yang}, {and} \bibinfo{person}{Qingyao Ai}.} \bibinfo{year}{2021}\natexlab{}.
\newblock \showarticletitle{ULTRA: an unbiased learning to rank algorithm toolbox}. In \bibinfo{booktitle}{\emph{CIKM'21}}. \bibinfo{pages}{4613--4622}.
\newblock


\bibitem[Wang et~al\mbox{.}(2021)]%
        {wang2021non}
\bibfield{author}{\bibinfo{person}{Nan Wang}, \bibinfo{person}{Zhen Qin}, \bibinfo{person}{Xuanhui Wang}, {and} \bibinfo{person}{Hongning Wang}.} \bibinfo{year}{2021}\natexlab{}.
\newblock \showarticletitle{Non-clicks mean irrelevant? propensity ratio scoring as a correction}. In \bibinfo{booktitle}{\emph{WSDM'21}}. \bibinfo{pages}{481--489}.
\newblock


\bibitem[Yu et~al\mbox{.}(2023)]%
        {yu2023cir}
\bibfield{author}{\bibinfo{person}{Lulu Yu}, \bibinfo{person}{Keping Bi}, \bibinfo{person}{Jiafeng Guo}, {and} \bibinfo{person}{Xueqi Cheng}.} \bibinfo{year}{2023}\natexlab{}.
\newblock \showarticletitle{CIR at the NTCIR-17 ULTRE-2 Task}.
\newblock \bibinfo{journal}{\emph{arXiv preprint arXiv:2310.11852}} (\bibinfo{year}{2023}).
\newblock


\bibitem[Zhang et~al\mbox{.}(2023)]%
        {zhang2023towards}
\bibfield{author}{\bibinfo{person}{Yunan Zhang}, \bibinfo{person}{Le Yan}, \bibinfo{person}{Zhen Qin}, \bibinfo{person}{Honglei Zhuang}, \bibinfo{person}{Jiaming Shen}, \bibinfo{person}{Xuanhui Wang}, \bibinfo{person}{Michael Bendersky}, {and} \bibinfo{person}{Marc Najork}.} \bibinfo{year}{2023}\natexlab{}.
\newblock \showarticletitle{Towards disentangling relevance and bias in unbiased learning to rank}. In \bibinfo{booktitle}{\emph{SIGKDD'23}}. \bibinfo{pages}{5618--5627}.
\newblock


\bibitem[Zou et~al\mbox{.}(2022)]%
        {zou2022large}
\bibfield{author}{\bibinfo{person}{Lixin Zou}, \bibinfo{person}{Haitao Mao}, \bibinfo{person}{Xiaokai Chu}, \bibinfo{person}{Jiliang Tang}, \bibinfo{person}{Wenwen Ye}, \bibinfo{person}{Shuaiqiang Wang}, {and} \bibinfo{person}{Dawei Yin}.} \bibinfo{year}{2022}\natexlab{}.
\newblock \showarticletitle{A large scale search dataset for unbiased learning to rank}.
\newblock \bibinfo{journal}{\emph{NeurIPS'22}}  \bibinfo{volume}{35} (\bibinfo{year}{2022}), \bibinfo{pages}{1127--1139}.
\newblock


\end{thebibliography}
